\documentclass[letterpaper,12pt]{article}
\usepackage{jheppub2} 
\usepackage[T1]{fontenc}
\usepackage{color,soul}
\usepackage{natbib}
\usepackage{graphicx}
\usepackage{hyperref}
\usepackage{amsmath}
\usepackage{amsfonts}
\usepackage{amssymb}
\usepackage{enumitem}
\usepackage{tocvsec2}
\usepackage{slashed}
\usepackage{comment}

\newcommand\beq{\begin{eqnarray}}
\newcommand\eeq{\end{eqnarray}}

\newcommand{\mybar}[1]%
        {\kern 0.6pt\overline{\kern -0.6pt#1\kern -0.6pt}\kern 0.6pt}
\DeclareMathOperator{\sech}{sech}
\newcommand{\norm}[1]{\lvert #1 \rvert}
\newcommand{\B}{\mathbf{B}}
\newcommand{\E}{\mathbf{E}}

\newcommand{\Br}{\mathbf{B}_\text{r}}
\newcommand{\Er}{\mathbf{E}_\text{r}}
\newcommand{\J}{\mathbf{J}}

\newcommand{\x}{\mathbf{x}}

\newcommand{\pt}{\partial_t}

\newcommand{\dd}{\text{d}}

\newcommand{\e}{\text{e}}

\newcommand{\A}{\mathbf{A}}

\newcommand{\rz}{\hat{\mathbf{x}}_z}
\newcommand{\sgn}{\text{sgn}}
\newcommand{\g}{\bigg(\frac{C\beta}{\pi f_a}\bigg)}

\hypersetup{
    colorlinks=true,   % false: boxed links; true: colored links
    linkcolor=red,     % color of internal links (box color = linkbordercolor)
    citecolor=blue,    % color of links to bibliography
    filecolor=magenta, % color of file links
    urlcolor=blue      % color of external links
}

\makeatletter
\def\l@subsubsection#1#2{}
\makeatother    

\advance\footnotesep 1.5pt

\begin{document}

\title{Electromagnetic radiation from axion condensates in a time dependent magnetic field}
\author[1]{Srimoyee Sen,}
\emailAdd{srimoyee08@gmail.com}
\author[1]{Lars Sivertsen,}
\emailAdd{lars@iastate.edu}
\affiliation[1]{Department of Physics and Astronomy,  Iowa State University, Ames IA 50011}

%\author[1]{Lars Sivertsen,}
%\emailAdd{lars@iastate.edu}
%\affiliation[1]{Department of Physics and Astronomy,  Iowa State University, Ames IA 50011}
\abstract{
Time dependent magnetic fields can be sourced by spinning neutron stars, orbiting binaries and merging neutron stars. We consider electromagnetic radiation from axion condensates in the background of an alternating magnetic field. We find that a resonant peak in radiation can occur when the frequency of the alternating magnetic field is comparable with the axion mass scale.  More interestingly, in situations where the frequency of the alternating magnetic field itself changes with time, as can be the case in binary mergers due to steady increase in orbital frequency, the resonant peak in radiation may occur for a range of axion mass scales scanned by the time-varying magnetic field frequency.  
} 

\maketitle

\maxtocdepth{subsection} 

\section{Introduction}
Axions were originally postulated to solve the strong CP problem \cite{PhysRevLett.38.1440, PhysRevLett.40.223} and are considered to be one of the prime dark matter candidates \cite{Preskill:1982cy, Abbott:1982af, Kim:1986ax, Cheng:1987gp, Raffelt:1990yz, Duffy:2009ig}. Owing to their Bose statistics, they can form coherently oscillating Bose-Einstein condensates (BECs) \cite{Visinelli:2017ooc, Kolb:1993hw, PhysRevLett.117.121801, Eby:2015hyx, Eby:2016cnq, PhysRevLett.118.011301, Helfer:2016ljl, Braaten:2016dlp, Bai:2016wpg, Eby:2017xaw, Schiappacasse:2017ham}. In theories with non-negligible axion-photon coupling these condensates can decay to photons via stimulated or spontaneous emission \cite{Ikeda:2018nhb, Sen:2018cjt, Boskovic:2018lkj, Boskovic:2019qao, Rosa:2017ury, Hertzberg:2018zte}. They can also decay by emitting electromagnetic radiation in the presence of an external magnetic field. The rate of depletion of the axion condensate in this case depends on the strength of the magnetic field, the frequency of the oscillation of the axion BEC and the axion photon coupling. There have been some recent studies of electromagnetic radiation from axion condensates in a static background magnetic field \cite{Amin:2021tnq, Dietrich:2018jov}. In this paper we take steps to understand how a time-varying magnetic field affects this radiation. More specifically we consider periodically varying magnetic fields (alternating magnetic fields) and investigate how the frequency of this periodic variation affects electromagnetic radiation. 

It is clear that if the frequency of the alternating external magnetic field is much smaller compared to the axion mass scale, then it is a good approximation to treat the magnetic field as static. However when the two scales are comparable we can expect the magnetic field frequency to leave its imprints on the radiation from the condensate. Some astrophysical scenarios where we find time varying magnetic field include spinning or orbiting stars, in particular neutron stars due to their strong magnetic fields. The magnetic field of a neutron star is typically dipolar and the axis of rotation of the star is often misaligned with respect to the axis of the dipole. As a result, a spinning neutron star can give rise to a rotating magnetic field which can be described as an alternating magnetic field with a frequency set by the spin rate of the star. Similarly, orbital motion of strongly magnetized neutron stars can also give rise to time varying magnetic fields in its environment. Note that periodic variation is not the most general form of time dependence found in neutron star environments. In fact,  neutron star mergers are speculated to give rise to exponentially growing magnetic fields,  growing from $10^{12}$ Gauss to $10^{18}$ Gauss in a matter of a few milliseconds \cite{Kiuchi:2015sga, Skoutnev:2021chg}. We do not analyze such rapidly growing magnetic fields in this paper and leave it for future work.  

In our analysis we assume the axion photon coupling to be weak which allows us to compute electromagnetic radiation from axion condensates perturbatively. We ignore the effect of axion back reaction on the condensate. This is justified so long as the back reaction time scale is larger compared to the time period of outgoing electromagnetic radiation. Along with an external alternating magnetic field we also take into account the effect of a background plasma on the radiation. 

Note that electromagnetic radiation from an axion condensate in the background of a spatially constant and static electromagnetic field was considered in ref \cite{Amin:2021tnq}. It was found that the radiated power is highly sensitive to the size of the axion condensate. More precisely, for an axion mass of $m_a$, the radiation from an axion condensate of size $R\gg \frac{1}{m_a}$ was found to be exponentially suppressed. For $m_a R\sim 1$ the condensates radiated efficiently. 
One interesting feature of this analysis was that the presence of a background plasma was found to increase electromagnetic radiation from large condensates with $R\gg \frac{1}{m_a}$. In fact, a resonant enhancement of the radiated power was obtained when the plasma frequency was very close to the axion mass scale. Our analysis in this paper demonstrates that a similar enhancement in radiated power can occur when the background magnetic field has a periodic variation in time irrespective of whether there exists a background plasma. In order for this resonance to take place, the frequency of the alternating field must be close to the axion mass scale. If we now introduce a background plasma with a plasma frequency of the order of the frequency of the alternating magnetic field, then the axion mass scale at which a resonant enhancement takes place depends on both the plasma frequency and the alternating magnetic field frequency. In this case, all three scales in the problem, namely the axion mass, the plasma frequency and the frequency of the alternating magnetic field are of the same order. In general , there is no reason for plasma frequency of a medium and the frequency of the alternating magnetic field to coincide. However, such a coincidence can be found if one compares the plasma frequency of the interstellar medium with the spin frequency of a pulsar. The interstellar medium has a plasma frequency of about a few kHz which is of the order of the spin frequency of fast spinning pulsars. Thus a condensate of ultralight axions of mass $10^{-12}$ eV (frequency $\sim$kHz) submerged in the interstellar medium (ISM) can emit significant electromagnetic radiation when it comes in contact with the alternating magnetic field of a pulsar. This serves as one of the motivations for taking into account the effects of both an alternating magnetic field and a plasma frequency on electromagnetic radiation from axion condensates. 

Note that, although resonant enhancement of radiation can take place in the absence of an alternating magnetic field when axion mass is of the order of the plasma mass of the medium, to realize this enhancement, the axion mass has to be very close to the plasma mass for large axion condensates. E.g. for an axion condensate of size $R\sim 100 m_a^{-1}$, the plasma mass must be within a few percent of the axion mass. This precise coincidence may not in general be realized even when the order of magnitude of the axion mass is the same as that of the plasma mass. However, when an alternating magnetic field is introduced, the resonant axion mass not only depends on the plasma mass, but also on the frequency of the magnetic field which itself can vary with time. \footnote{For example, in a binary neutron star merger, the orbital frequency of the merging objects can increase. Similarly, accretion can increase the rate of spin of a star contributing to an increasing frequency of rotation for the magnetic field.} As a result, a changing frequency of the magnetic field changes the resonance condition: the axion mass scale which can experience resonant enhancement in radiation changes with time. Therefore, even if the axion mass does not exactly coincide with the resonant mass at a certain instant in time, it can do so at a later instant due to the time variation of the resonant mass. Thus, condensates of a range of mass scales can decay through radiation in the presence of an alternating background magnetic field of time dependent frequency. % provided the time variation of the frequency is not too fast. 

Motivated by the above considerations, we take the first step towards understanding the effect of alternating magnetic fields on electromagnetic radiation from axion condensate in this paper. Note that the spatial dependence of magnetic fields in astrophysical environment can be complicated in general. We do not take into account such complexities here. Instead we model the external field as a spatially constant background magnetic field which alternates in time. We however do take into account the finite extent of the axion condensate. We leave the analysis of a spatially inhomogeneous magnetic field for future work. 

The organization of the paper is as follows: We begin with a review of electromagnetic radiation from axion condensates in the presence of a 
spatially constant time independent magnetic field. We then discuss the resonant enhancement in radiation when the external magnetic field has a periodic time variation. This analysis is followed by a discussion of how a background plasma and an alternating magnetic field together affect the resonance in radiation. In the next part of the paper we introduce an external alternating magnetic field with time varying frequency and 
demonstrate how a large axion condensate which was not radiating at a certain instant in time begins to radiate efficiently at a later instant. We conclude with a section on how resonant enhancement in radiation affects axion decay (back reaction) time scale. 

\label{sec1}
%\section{Radiated Power}
\section{Electromagnetic radiation from axion consates in a background magnetic field}
%In this section we will begin with a discussion of axion-photon interactions and how an axion BEC radiates in the presence of a constant background electromagnetic field. We will then introduce a background magnetic field alternating in time and analyse the corresponding radiation
%for a constant frequency and then for a time varying frequency for the external field. 
%In this section, our goal is to compute the electromag

The axion-photon Lagrangian of interest to us is given by
\begin{equation}
\mathcal{L} = -\frac{1}{4}F^{\mu\nu}F_{\mu\nu}+J_m^{\mu}A_{\mu} +\frac{C\beta}{4\pi f_a}\phi\epsilon^{\mu\nu\lambda\rho} F_{\mu\nu}F_{\lambda\rho}+\frac{1}{2}(\partial_\mu\phi)(\partial^\mu\phi)-\frac{1}{2}m_a^2\phi^2-V(\phi)+\cdots.
\end{equation}
Here $\phi$ is the pseudo-scalar axion field, $m_a$ is the mass of the axion, $V(\phi)$ includes self interaction of the axion field and $\beta$ is the fine-structure constant. In the axion-photon interaction term, $C$ is a model dependent number and $f_a$ is the axion decay constant. We have also included in the Lagrangian a coupling between the electromagnetic field $A_{\mu}$ and matter fields via a current density $J_m^{\mu}$. This operator will describe the physics of any background plasma around the axion condensate as well as any matter source generating external electromagnetic fields. In the absence of any electromagnetic coupling ($\beta=0$), local gravitational interactions combined with axion self interaction $V(\phi)$ can give rise to a stable axion BEC: a coherently oscillating axion condensate with frequency $\sim m_a$ \cite{Schiappacasse:2017ham}. These axion BECs have fixed particle numbers.  When electromagnetic interactions are turned on, particle number can change and it may be possible for the axion BEC to decay to photons. In the absence of any external electromagnetic field there are two mechanisms via which axions can decay to photons: spontaneous emission and stimulated emission \cite{Sen:2018cjt}. The spontaneous emission rate for axions go as $\frac{m_a^3}{f_a^2}$ and for light axions this rate is minuscule. For example, for a QCD axion ($\Lambda_{\text{QCD}}\sim 100$ MeV) of mass $10^{-12}$ eV, which corresponds to a frequency scale of about $1$ kHz, the spontaneous emission rate is about $10^{-88}s^{-1}$ and the corresponding decay time scale is much larger than the age of the universe making this process irrelevant for time scales of interest to us. The rate of stimulated emission for a homogeneous condensate on the other is given by $\frac{C\beta|\tilde{\phi}|}{\pi f_a}m_a$ where $|\tilde{\phi}|$ is the amplitude for the oscillation of the axion field. This emission rate can be much larger than the rate of spontaneous emission. For example, with $\frac{C\beta|\tilde{\phi}|}{\pi f_a}\sim 1$, for a homogeneous QCD axion condensate of axion mass $m_a \sim 10^{-12}$eV, the decay time scale is of the order of $10^{-3}$s. Note however, for a BEC of a finite size, the rate of decay can be very sensitive to the size of the condensate. It was shown in \cite{Hertzberg:2018zte} that stimulated emission can only take place for $\frac{C\beta|\tilde{\phi}|m_a}{\pi f_a}> \frac{1}{R}$ where $R$ was the size of the condensate. %Thus, in the limit of weak coupling, a condensate of size $R\sim\frac{1}{m_a}$ is unlikely to disappear via stimulated emission.
We can therefore imagine coherently oscillating axion condensates having a long life-time despite their coupling to electromagnetic fields, as long as there are no external electromagnetic fields present and the axion-photon coupling is not too strong.

With this we can now proceed to understand the response of axion BECs to external electromagnetic fields. Since we are specifically interested in external magnetic fields we will set external electric fields to zero. Our analysis can however be extended to include external electric fields in a rather straightforward manner. 
\subsection{Radiation in a constant magnetic field}
To understand the response of axion BECs to electromagnetic fields we first write down the axion-photon equation of motion 
\begin{align}
\nabla \times \B(\x,t)-\pt \E(\x,t)-\J_m(\x,t) &= -\frac{C\beta}{\pi f_a}\Big[\big(\pt\phi(\x,t)\big)\B(\x,t)+\nabla\phi(\x,t)\times \E(\x,t)\Big],\\
\nabla\times\E(\x,t) &= -\pt\B(\x,t),\\
\nabla\cdot\E(\x,t) &= \rho_m(\x,t)+\frac{C\beta}{\pi f_a}\nabla\phi(\x,t)\cdot\B(\x,t),\\
\nabla\cdot\B(\x,t) &=0,
\label{Max}
\end{align}
where $\J_m$ and $\rho_m$ are the current and charge density corresponding to some matter fields interacting with the electromagnetic field. We can divide up the matter four-current into two contributions, $J_m^{\mu}=J_s^{\mu}+J_p^{\mu}$ where $J_s^{\mu}$ will describe any matter current sourcing background electromagnetic fields and $J_p^{\mu}$ will describe the physics of any plasma medium if present. In the absence of an electromagnetic interaction, coherently oscillating axion BECs of frequency very close to the axion mass can form owing to their gravitational interaction and the axion self interaction $V(\phi)$ \cite{Schiappacasse:2017ham}. These interactions also determine the spatial profile for these axion BECs. We don't discuss the details of the derivation of these spatial profiles and the stability criterion of these BECs in this paper. We will instead use the profile of the axion condensate found in \cite{Schiappacasse:2017ham} to understand the response of axions to background magnetic fields. We focus on spherically symmetric axion clumps in this paper which can be described with the following ansatz
\beq
\phi(\x,t)=\tilde{\phi}\cos(\mu t)\sech\Big[\frac{\norm{\x}}{R}\Big],
\label{ansatz}
\eeq
where $R$ is the radius of the axion condensate and $\mu$ is the oscillation frequency of the axion condensate with $\mu\sim m_a$.  
With this ansatz we can now proceed to analyze the physics of a coherently oscillating axion condensate in an external magnetic field. We will first review the radiation from axion condensates in the background of a spatially constant time independent magnetic field which will closely follow the analysis in \cite{Amin:2021tnq}. After this discussion we will move on to discussing alternating time dependent magnetic fields. 

In the presence of an external field, Maxwell's equation as written in Eq. \ref{Max} will effectively include an oscillating current source due to the coupling of axion condensate with the background field. This alternating current source in turn produces electromagnetic radiation which can take away energy from the axion condensate. To elaborate on this, let us write the electromagnetic fields as
$\bf{E}=\bf{E}_0+\bf{E}_r$ and $\bf{B}=\bf{B}_0+\bf{B}_r$. Here $\bf{E}_0$ and $\bf{B}_0$ correspond to external and $\bf{E}_r$ and $\bf{B}_r$ to radiated electromagnetic fields. Writing the matter current as $J^{\mu}_m=J_p^{\mu}+J_s^{\mu}$ we can re-express Maxwell's equations as
\begin{align}
&\nabla \times \B_0(\x,t)-\pt \E_0(\x,t)-\J_s(\x,t) =0,\nonumber\\
&\nabla\times\E_0(\x,t) = -\pt\B_0(\x,t),\nonumber\\
&\nabla\cdot\E_0(\x,t) = \rho_s(\x,t),\nonumber\\
&\nabla\cdot\B_0(\x,t) =0,
\label{Max2}
\end{align}
and 
\begin{align}
\nabla \times \B_r(\x,t)-\pt \E_r(\x,t)-\J_p(\x,t) &= -\frac{C\beta}{\pi f_a}\Big[\big(\pt\phi(\x,t)\big)\B_0(\x,t)+\nabla\phi(\x,t)\times \E_0(\x,t)\Big],\nonumber\\
\nabla\times\E_r(\x,t) &= -\pt\B_r(\x,t),\nonumber\\
\nabla\cdot\E_r(\x,t) &= \rho_p(\x,t)+\frac{C\beta}{\pi f_a}\nabla\phi(\x,t)\cdot\B_0(\x,t),\nonumber\\
\nabla\cdot\B_r(\x,t) &=0.
\label{Max3}
\end{align}
Eq. \ref{Max2} describes the the external electric and magnetic field sourced by some matter current $J^{\mu}_s$, the details of which are uninteresting for the problem at hand. Eq. \ref{Max3} describes the radiated electromagnetic fields and how this radiation depends on the 
background electromagnetic fields $\bf{E}_0$ and $\bf{B}_0$, the axion condensate and the matter current of the medium $J_p^{\mu}$.
In writing Eq. \ref{Max3} we have assumed that the radiated electromagnetic fields are small compared to the background electric and magnetic fields, i.e. $\bf{E}_r, \bf{B}_r \ll \bf{E}_0, \bf{B}_0$. This is justified as long as the axion-photon coupling is small so that a weak coupling analysis holds. As is clear from Eq. \ref{Max3}, the axion couplings to external electric and magnetic fields act like an oscillating current source for the radiated fields $\bf{E}_r$ and $\bf{B}_r$.  
If we set the matter current of the medium $J_p^{\mu}$ and the external electric field $\bf{E}_0$ to zero, we can solve Eq. \ref{Max3} with a current and charge density 
\beq
\J(\x,t)&=&-\frac{C\beta}{\pi f_a}\Big[\big(\pt\phi(\x,t)\big)\B_0(\x,t)\Big]\sim\frac{C\beta}{\pi f_a}m_a\tilde{\phi}\sin(m_a t)\sech\Big[\frac{\norm{\x}}{R}\Big]B_0\rz,\nonumber\\
\rho(\x,t)&=&\frac{C\beta}{\pi f_a}\nabla\phi(\x,t)\cdot\B_0(\x,t)\sim- \frac{C\beta}{\pi f_a}\tilde{\phi}\cos(m_a t)\frac{\sech\big[\frac{\norm{\x}}{R}\big]\tanh\big[\frac{\norm{\x}}{R}\big]}{R}B_0\,\,\hat{\x}\cdot\hat{\x}_z,\nonumber\\
\eeq 
  where we have substituted $\B_0(x,t)$ with a constant magnetic field in the $z$ direction $\B(\x,t)=B_0\rz$. 
	
Using the standard retarded Green's function 
\beq
G(x,t;x't')=-\frac{\delta(t-t'-|\x-\x'|)}{4\pi|\x-\x'|},
\eeq 
we can now obtain the radiated electric and magnetic fields in the radiation zone (at a distance much larger than the size of the axion condensate)
\begin{align}
\Er(\x,t) &= -\pt \A(\x,t)-\nabla A^0(\x,t)
\nonumber
\\
&\approx\frac{C\beta}{\pi f_a}\frac{\tilde{\phi}B_0m_a\pi^2 R^2}{4\norm{\x}}\bigg[\frac{\tanh(\pi k_{m_a} R/2)}{\cosh(\pi k_{m_a} R/2)}\cos(m_a t-k_{m_a}\norm{\x})\bigg]\rz
\nonumber
\\
&+\g \frac{\tilde{\phi} B_0 k_{m_a} \pi^2 R^2}{4\norm{\x}}\bigg[\frac{\tanh(\pi k_{m_a}R/2)}{\cosh(\pi k_{m_a}R/2)}\cos(m_a t- k_{m_a}\norm{\x})
\bigg]\hat{\x},
\nonumber
\\
\Br(\x,t) &=\nabla\times\A(\x,t)
\nonumber
\\
&\approx \frac{C\beta}{\pi f_a}\frac{\tilde{\phi}B_0m_a\pi^2 R^2}{4\norm{\x}}\bigg[\frac{\tanh(\pi k_{m_a} R/2)}{\cosh(\pi k_{m_a} R/2)}\cos(m_a t-\norm{\x}k_{m_a})\bigg](\hat{\x}\times\rz),\label{225}
\end{align} 
where $k_{m_a}=m_a$.
The corresponding time-averaged radiated power is given by
\beq
P=\left(\frac{C\beta}{\pi f_a}\right)^2\frac{\tilde{\phi}^2B_0^2m_a^2R^4\pi^5}{12}\left(\frac{\tanh(\pi k_{m_a} R/2)}{\cosh(\pi k_{m_a} R/2)}\right)^2.
\label{power}
\eeq
Note that even though $\mathbf{E}_r=\partial_t \mathbf{A}-\nabla A_0$, $\nabla A_0$ does not contribute to the outgoing radiated power. Thus, if we define $\bf{E}_r'=\mathbf{E}_r+\nabla A_0$ and $\mathbf{B}_r'=\mathbf{B}_r$, the outgoing radiated power can be re-expressed as $|\mathbf{x}|^2\int d\omega\,\hat{\x}\cdot\left(\bf{E}_r'\times\mathbf{B}_r'\right)$ where $\omega$ denotes the solid angle.
It is clear from this expression that the power radiated from axion condensates peaks when the radius of the clump is of the order of the inverse axion mass. For condensates that are much larger than the inverse mass scale of the axion, radiation is exponentially suppressed. It was shown in \cite{Amin:2021tnq} that upon introduction of a background plasma, this exponential suppression can be eliminated when the plasma frequency is very close to the axion mass scale. In that case, it may be possible for axion condensates of sizes much larger than the inverse axion mass scales to radiate efficiently. In the next section we will see that a similar enhancement can be achieved with an alternating background magnetic field even in the absence of a plasma.% In the next section we analyze the effect of an alternating magnetic field on radiation from axion condensate. We will find that when the frequency of the alternating magnetic field coincides with the axion mass, there can be resonant enhancement of radiation which enables large axion condensates to radiate. Eventually we will also include the effect of a background plasma along with an alternating magnetic field to understand the combined effect of the plasma and the time variation of the magnetic field.  
\subsection{Radiation in an alternating magnetic field}
We will begin with an external background field which is alternating in time, but is spatially constant. We plan to relax this restriction in future work. Since the purpose of this analysis is to understand the response of axion condensates to the time-dependence of the external field specifically, we refrain from introducing more complex spatial dependence in the background field profile. In order to proceed, in Eq. \ref{Max3} we assume that the contribution to the current density from the coupling between any external electric field and the axion condensate is much smaller than the contribution to the current density from the magnetic field coupling to axions. Thus we work in the limit 
\beq
\nabla\phi\times\bf{E}_0\ll(\partial_t\phi) \mathbf{B}_0.
\eeq
The magnetic field is chosen to be of the form 
\beq
\B_0=B_0 \cos(\Omega t+\gamma)\rz,
\eeq
where $\gamma$ is some arbitrary phase shift. We also include in the axion ansatz an arbitrary phase shift of the form 
\beq
\phi=\tilde{\phi}\cos(m_a t+\alpha)\sech\Big[\frac{\norm{\x}}{R}\Big],
\eeq
in order to explore how a phase difference between the axion field and the external magnetic field affects outgoing radiation. 
The corresponding current density in Maxwell's equation can be written as
\beq
\J(\x,t)&=&-\frac{C\beta}{\pi f_a}\Big[\big(\pt\phi(\x,t)\big)\B_0(\x,t)\Big]\nonumber\\
&\sim&\left(\frac{C\beta}{\pi f_a}m_a\tilde{\phi}\right)\left(\frac{\sin((m_a+\Omega) t+\alpha+\gamma)+\sin((m_a-\Omega) t+\alpha-\gamma)}{2}\right)\sech\Big[\frac{\norm{\x}}{R}\Big]B_0\rz\nonumber\\
\\
\rho(\x,t)&=&\frac{C\beta}{\pi f_a}\nabla\phi(\x,t)\cdot\B_0(\x,t)\nonumber\\
&\sim& -\left(\frac{C\beta}{\pi f_a}\tilde{\phi}\right)\left(\frac{\cos((m_a+\Omega) t+\alpha+\gamma)+\cos((m_a-\Omega) t+\alpha-\gamma)}{2}\right)
\nonumber
\\
&&\quad\quad\quad\quad\quad\quad\times\left(\frac{\sech[\frac{\norm{\x}}{R}]\tanh[\frac{\norm{\x}}{R}]}{R}\right)B_0\, \hat{\x}\cdot\hat{\x}_z.\nonumber\\
\eeq 
Without a loss of generality we can take $\Omega>0$. Note that when we compute the radiated power in outgoing waves we will have to be careful in taking into account both $m_a>\Omega$ or $m_a<\Omega$. The corresponding retarded Green's function is given by
\beq
G(xt;x't')&=&-\int \frac{d\omega}{2\pi}\frac{1}{4\pi|\vec{x}-\vec{x}'|}\left(e^{i|\omega||\vec{x}-\vec{x}'|-i\omega(t-t')}\theta(\omega)+e^{-i|\omega||\vec{x}-\vec{x}'|-i\omega(t-t')}\theta(-\omega)\right)\nonumber\\
\eeq
which can be used to find the radiated electric and magnetic fields
\beq
\mathbf{E}'_\text{r}&\approx&\left(\frac{C\beta}{\pi f_a}\frac{B_0 \tilde{\phi}m_a}{8|x|}R^2\pi^2\right)\left[\frac{\tanh\left(\frac{\pi(m_a+\Omega)R}{2}\right)}{\cosh\left(\frac{\pi(m_a+\Omega)R}{2}\right)}\cos((m_a+\Omega)(t-|\bf{x}|)+\alpha+\gamma)\rz \right.\nonumber\\
&&\left.+ \sgn(m_a-\Omega)\frac{\tanh\left(\frac{\pi|(m_a-\Omega)|R}{2}\right)}{\cosh\left(\frac{\pi|(m_a-\Omega)|R}{2}\right)}\cos(|(m_a-\Omega)|(t-|{\bf x}|)+\sgn(m_a-\Omega)(\alpha-\gamma))\rz\right],\nonumber\\
\eeq
and 
\beq
\mathbf{B}_\text{r}&\approx&(\hat{\x}\times\rz)\left(\frac{C\beta}{\pi f_a}\frac{B_0 m_a\tilde{\phi}}{8|x|}R^2\pi^2\right)\left[\frac{\tanh\left(\frac{\pi(m_a+\Omega)R}{2}\right)}{\cosh\left(\frac{\pi(m_a+\Omega)R}{2}\right)}\cos((m_a+\Omega)(t-|\bf{x}|)+\alpha+\gamma)\right.\nonumber\\
&&\left.+ \sgn(m_a-\Omega)\frac{\tanh\left(\frac{\pi|(m_a-\Omega)|R}{2}\right)}{\cosh\left(\frac{\pi|(m_a-\Omega)|R}{2}\right)}\cos(|(m_a-\Omega)|(t-|{\bf x}|)+\sgn(m_a-\Omega)(\alpha-\gamma))\right].\nonumber\\
\eeq
Here the $\sgn(\sigma)$ stands for the sign of the variable $\sigma$. The approximate signs are there to remind us that we have ignored $\sim |\bf{x}|^{-n}$ terms where $n>1$ since they don't contribute to radiation. 
%\hl{Note that we have omitted terms that are proportional to $\sim |\bf{x}|^{-3}$ as these will be negligible far away from the source. For $m_a<\Omega$, the power radiated per solid angle, $\Omega_{\text{solid}}$, can be written as}  \textcolor{red}{(SENTENCE CHANGED FROM "the magnitude of the corresponding Poynting vector can be written as")}
For $m_a<\Omega$, the instantaneous power radiated is given by
\beq
P_i&=&\frac{4\pi}{3}\left(\frac{C\beta}{\pi f_a}\frac{B_0 m_a \tilde{\phi}}{8}\pi^2R^2\right)^2\left[
\frac{\tanh\left(\frac{\pi(m_a+\Omega)R}{2}\right)^2}{\cosh\left(\frac{\pi(m_a+\Omega)R}{2}\right)^2}\left(1+\cos(2(m_a+\Omega)(t-|\mathbf{x}|)+2\alpha+2\gamma)\right)\right.\nonumber\\
&&\left.-2\frac{\tanh\left(\frac{\pi(m_a+\Omega)R}{2}\right)}{\cosh\left(\frac{\pi(m_a+\Omega)R}{2}\right)}\frac{\tanh\left(\frac{\pi|m_a-\Omega|R}{2}\right)}{\cosh\left(\frac{\pi|m_a-\Omega|R}{2}\right)}\left(\cos(2\Omega(t-|\mathbf{x}|)+2\gamma)+\cos(2m_a(t-|\mathbf{x}|)+2\alpha)\right)\right.\nonumber\\
&&\left.+\frac{\tanh\left(\frac{\pi|m_a-\Omega|R}{2}\right)^2}{\cosh\left(\frac{\pi|m_a-\Omega|R}{2}\right)^2}\left(1+\cos(2|m_a-\Omega|(t-|\mathbf{x}|)-2(\alpha-\gamma))\right)
\right],\nonumber\\
\label{power1}
\eeq
where the subscript $i$ stands for instantaneous. 
%\hl{where $\theta$ is the angle between $\hat{\bf{x}}$ and $\hat{\bf{z}}$. For $m_a>\Omega$ the power per solid angle is given by} \textcolor{red}{(Some factors are corrected)}
For $m_a>\Omega$ the instantaneous radiated power is
\beq
P_i&=&\frac{4\pi}{3}\left(\frac{C\beta}{\pi f_a}\frac{B_0 m_a \tilde{\phi}}{8}\pi^2R^2\right)^2\left[
\frac{\tanh\left(\frac{\pi(m_a+\Omega)R}{2}\right)^2}{\cosh\left(\frac{\pi(m_a+\Omega)R}{2}\right)^2}\left(1+\cos(2(m_a+\Omega)(t-|\mathbf{x}|)+2\alpha+2\gamma)\right)\right.\nonumber\\
&&\left.+2\frac{\tanh\left(\frac{\pi(m_a+\Omega)R}{2}\right)}{\cosh\left(\frac{\pi(m_a+\Omega)R}{2}\right)}\frac{\tanh\left(\frac{\pi|m_a-\Omega|R}{2}\right)}{\cosh\left(\frac{\pi|m_a-\Omega|R}{2}\right)}\left(\cos(2\Omega(t-|\mathbf{x}|)+2\gamma)+\cos(2m_a(t-|\mathbf{x}|)+2\alpha)\right)\right.\nonumber\\
&&\left.+\frac{\tanh\left(\frac{\pi|m_a-\Omega|R}{2}\right)^2}{\cosh\left(\frac{\pi|m_a-\Omega|R}{2}\right)^2}\left(1+\cos(2|m_a-\Omega|(t-|\mathbf{x}|)+2(\alpha-\gamma))\right)
\right].\nonumber\\
\label{power2}
\eeq
There are quite a few lessons we can learn from the expression for the radiated power. To begin with, one can concentrate on the two limits of $\Omega\gg m_a$ and $m_a\gg\Omega$. In the former case the expression for power radiated obtained from Eq. \ref{power1} should be recoverable by ignoring the axion frequency in the sinusoidal variation of the current density, i.e. by writing
\beq
\J(\x,t)&=&-\frac{C\beta}{\pi f_a}\left(\partial_t\phi\right) \mathbf{B}_0\nonumber\\
&=&\frac{C\beta}{\pi f_a}\tilde{\phi}\sin(\alpha)\sech\Big[\frac{\norm{\x}}{R}\Big] B_0\cos(\Omega t+\gamma)\rz.
\label{app1}
\eeq
This amount to substituting the axion field $\tilde{\phi}$ by $\tilde{\phi}\sin(\alpha)$ in Eq. \ref{power}.
Similarly, for the latter case we should be able to ignore the magnetic field frequency in the current density and the radiated power should be obtainable from replacing the magnetic field $B_0$ by $B_0\cos(\gamma)$ in Eq. \ref{power}. In order to see how these expression for radiated power emerge from Eq. \ref{power1}, let's consider the two limits $m_a\ll\Omega$ and $m_a\gg\Omega$ sequentially. For $m_a\ll\Omega$, in the expression for instantaneous power in Eq. \ref{power1} we should set $\frac{m_a}{\Omega}\rightarrow 0$ and then average the instantaneous power over the time scale of $2\pi/\Omega$. This produces the expected result for the limit of $m_a\ll \Omega$ obtained by substituting $\tilde{\phi}$ by $\tilde{\phi}\sin(\alpha)$ in Eq. \ref{power}. In the case of $\Omega\ll m_a$, one should replace $\Omega\rightarrow 0$ in the expression for the instantaneous power in Eq. \ref{power2},  and then average the instantaneous power over a period of time $2\pi/m_a$. This produces the expected result for the radiated power in the limit of $\Omega\ll m_a$ obtained from Eq. \ref{power} by substituting $B_0$ by $B_0\cos(\gamma)$.

Finally, if we are interested in the limit where $m_a$ and $\Omega$ are not very different, we can average the expression for the power radiated in Eq. \ref{power1} and \ref{power2} over a period of time much larger than either $\Omega$ or $m_a$ to obtain the time averaged radiated power $P_{\text{av}}$. We find that the time average radiated power in both cases, $m_a>\Omega$ and $\Omega>m_a$ are approximately the same and given by 
\beq
P_\text{av}&\approx&\frac{4\pi}{3}\left(\frac{C\beta}{\pi f_a}\frac{B_0 m_a \tilde{\phi}}{8}\pi^2R^2\right)^2\left[
\frac{\tanh\left(\frac{\pi(m_a+\Omega)R}{2}\right)^2}{\cosh\left(\frac{\pi(m_a+\Omega)R}{2}\right)^2}+\frac{\tanh\left(\frac{\pi|m_a-\Omega|R}{2}\right)^2}{\cosh\left(\frac{\pi|m_a-\Omega|R}{2}\right)^2}
\right].\nonumber\\
\label{power3}
\eeq 
The approximate sign in \ref{power3} takes into account the fact that the radiated power has contributions from electromagnetic waves of two different frequencies $|m_a-\Omega|$ and $m_a+\Omega$ and the time average computed in Eq. \ref{power3} is taken over a time interval longer than the longest time period of the two frequency scales, i.e. $\frac{2\pi}{|m_a-\Omega|}$. 
As we observe from Eq. \ref{power3} and the discussion preceding it, the average radiated power is independent of the initial phase shift of the magnetic field and the axion condensate unless $\frac{\Omega}{m_a}\ll 1$ or $\frac{m_a}{\Omega}\ll 1$. The dependence of radiation on the phase shift in the limits $m_a\ll \Omega$ and $\Omega\ll m_a$ can however be reinterpreted as a dependence on the amplitude of an approximately time independent axion condensate or a magnetic field respectively.

We see from Eq. \ref{power3} that radiated power depends sensitively on the size of the condensate as before. However, as opposed to the case in Eq. \ref{power} when $\Omega=0$, the peak in radiation occurs for two different values of the condensate size: one at $R\sim\frac{1}{|m_a-\Omega|}$, the other at $R\sim\frac{1}{m_a+\Omega}$. 
%In cases where $m_a\ll\Omega$ or $\Omega\ll m_a$, the two peaks approach each other and the radiated power peaks for condensate sizes that scale inversely with the larger frequency scale, i.e. $\frac{1}{m_a}$ in the former case and $\frac{1}{\Omega}$ in the latter. %For condensates that are much larger in size compared to either $\frac{1}{m_a+\Omega}$, radiation is exponentially suppressed.
When $m_a\sim\Omega$, if $\frac{|m_a-\Omega|}{m_a}\ll 1$, these two length scales can be very different. In this case, one of the peaks in radiation lies at a radius of the order of $\frac{1}{m_a+\Omega}\sim\frac{1}{m_a}\sim\frac{1}{\Omega}$. The other peak is at a much larger radius of the order of $\frac{1}{m_a-\Omega}\gg\frac{1}{m_a}\sim\frac{1}{\Omega}$. Thus, we find that there is a resonant enhancement of radiation for large axion condensates when the frequency of the oscillating magnetic field coincides with the axion mass. This enhancement is similar to the resonant enhancement found in \cite{Amin:2021tnq} in the presence of a background plasma. %We postpone the discussion of a finite plasma frequency to the section \ref{plasma}. %
In the next section we will introduce a background plasma medium and analyze its effects on radiation combined with the effects of an alternating magnetic field. %In order to see how this comes about note the two expressions for the radiated power in Eq. \ref{power1} and \ref{power2}. When $\frac{|m_a-\omega|}{m_a}\ll 1$, the first term on the RHS of Eq. \ref{power1} and \ref{power2} peaks when $R\sim \frac{1}{m_a+\Omega}$ and the other terms remain smaller. On the other hand, for condensated of size $R\sim\frac{1}{|m_a-\Omega|}$, it is the last term on the RHS that dominates over the other two giving rise to a second resonant peak in axion decay. 
\subsection{Effects of a background plasma}
\label{plasma}
%Explain Drude model, show resonance, explain why even though $k \sim \sqrt{\omega_{\pm}^2-\omega_p^2}\ll \omega_p$ we can still have decay for the axion: it is because $\frac{\omega_p^2}{(\omega^2-\omega_p^2)}\frac{1}{\omega_p\tau}$ is still really small. So, even though the signal may not reach us, it will dissipate in interstellar plasma. 

%We will now introduce a background plasma and analyze its effects on axion decay. Note that, 
The effects of a background plasma on radiation can be modeled effectively by replacing the matter current of the medium $J_p$ by $\sigma E$ where $\sigma$ is the conductivity of the plasma. If we are interested in an electron ion plasma as in the case of the interstellar medium, $\sigma$ has a frequency dependence which can be modeled using the Drude model as 
\beq
\sigma(\omega)=\frac{\sigma_0}{1-i\omega \tau}. 
\eeq
Here $\tau$ is the collision time for electrons. In the collision less limit, where the collision frequency is much smaller than the frequency of the electromagnetic radiation, i.e. $\omega\tau\gg 1$, we can write the conductivity as 
\beq
\sigma\approx i\frac{\sigma_0}{\omega\tau}.
\eeq
If we substitute this expression for conductivity in Maxwell's equations, we find that the photon acquires a plasma mass $\omega_P=\sqrt{\frac{\sigma_0}{\tau}}$ with a dispersion relation of the form $\omega^2=k^2+\omega_P^2$. The conductivity $\sigma_0$ can be expressed as $\sigma_0=\frac{4\pi n_e e^2\tau}{m_e}$ where $m_e$ is the electron mass, $e$ is the electron's charge and $n_e$ is the density of electrons. Thus, the plasma mass is given by $\omega_P=\sqrt{\frac{4\pi n_e e^2}{m_e}}$. As shown in \cite{Amin:2021tnq}, the electromagnetic radiation emitted from an axion condensate in the presence of a constant external magnetic field and a plasma in the collision-less limit is given by 
\beq
P_{\text{av}}=\left(\frac{C\beta}{\pi f_a}\right)^2\frac{\tilde{\phi}^2B_0^2m_a k_{m_a}R^4\pi^5}{12}\left(\frac{\tanh(\pi k_{m_a} R/2)}{\cosh(\pi k_{m_a} R/2)}\right)^2
\label{power31}
\eeq
 with $k_{m_a}=\sqrt{m_a^2-\omega_P^2}$. This expression also makes it clear how a plasma can enhance electromagnetic radiation from a large axion condensate. The radiated power is dependent on the size of the condensate such that the radiation is peaked for condensates of size $R\sim\frac{1}{k_{m_a}}$ and exponentially suppressed otherwise. When $\omega_P$ is negligible or even of the order of $m_a$, the peak in radiation generally occurs for axion condensates of size $\frac{1}{k_{m_a}}\sim\frac{1}{m_a}$. 
When $\frac{\sqrt{m_a^2-\omega_P^2}}{m_a}\ll 1$, however, condensates of much larger size than $\frac{1}{m_a}$ can radiate efficiently. More precisely, radiation peaks when the condensate radius $R$, the axion mass $m_a$, and the plasma frequency $\omega_\text{P}$ are related by $R=\frac{\frac{2}{\pi}\log(\sqrt{2}+1)}{\sqrt{m_a^2-\omega_P^2}}$. Thus, the larger the condensate, the closer the axion mass and the plasma frequency have to be in order for enhanced radiation to take place.

In general though, there is no reason to expect the axion mass and the plasma frequency to precisely coincide even if they are of the same order of magnitude. Thus we don't expect the resonance in radiation to take place for very large axion condensates. 
However, in the presence of an alternating background magnetic field with time varying frequency, it may be possible to achieve this resonant conversion even when plasma frequency and axion mass don't exactly coincide. This is because the axion mass suited for resonant conversion should depend both on the plasma frequency and the frequency of the alternating magnetic field. Thus, with an alternating magnetic field whose frequency changes with time, the condition for resonant radiation can be met over a certain interval of time as the frequency changes.
%When the frequency of the magnetic field itself is time dependent, we can expect the axion mass suited for resonant conversion change with time. 
So long as the plasma frequency, the mass of the axion and the frequency of the alternating magnetic field are of the same order of magnitude, a time varying frequency for the magnetic field is likely to scan a range of axion masses for which enhanced or resonant radiation can take place. % Thus, the condition for resonance will be achieved at a certain instant in time making the resonant decay of the axions possible.
It is therefore useful to analyze the combined effects of a plasma background and an alternating magnetic field on the radiation from axion condensate as we will do next. % We will now consider axion condensates in the background of an alternating magnetic field in the presence of a background plasma. 
In our analysis here, we will treat the frequency of the alternating magnetic field $\Omega$ to be a constant in time. This analysis should correctly model a resonant enhancement in radiation even for time varying $\Omega$ as long as the time variation of $\Omega$ over a single time period of oscillation of the radiated field remains small compared to the frequency of the radiated field. %This will help us understand the condition for resonant enhancement 
%the effect of an alternating magnetic field on radiation from an axion condensate in the background of a plasma.
 %Note that, to understand the resonance condition for 
%axions in an alternating magnetic field and in a background plasma, it is sufficient to consider the alternating magnetic field to have a constant frequency $\Omega$. This is because, even if we are interested in a time varying $\Omega$, we only need the time variation to alter $\Omega$ slowly so as to realize the resonance condition at a certain instant in time. Also, in order for axion condensate to be able to decay via resonance we will want $\Omega$ to not vary too rapidly: $\frac{\dot{\Omega}}{\Omega}\ll \Omega$.
In this case, again to account for both of the possibilities $\Omega>m_a$ and $m_a>\Omega$ we can write the retarded Green's function as
\beq
G(xt;x't')&=&-\int \frac{d\omega}{2\pi}\frac{\theta\left(\omega^2-\omega_p^2\right)}{4\pi|\vec{x}-\vec{x}'|}\left(e^{i\sqrt{\omega^2-\omega_p^2}|\vec{x}-\vec{x}'|-i\omega(t-t')}\theta(\omega)+e^{-i\sqrt{\omega^2-\omega_p^2}|\vec{x}-\vec{x}'|-i\omega(t-t')}\theta(-\omega)\right)\nonumber\\
&-&\int \frac{d\omega}{2\pi}\frac{\theta\left(-\omega^2+\omega_p^2\right)}{4\pi|\vec{x}-\vec{x}'|}e^{-\sqrt{|\omega^2-\omega_p^2|}|\vec{x}-\vec{x}'|-i\omega(t-t')}.
\eeq
Here we have also taken into account of the fact that radiation for frequencies smaller than the plasma frequency is exponentially damped. We can now write down the time averaged radiated power 
\beq
P_\text{av}&\approx&\frac{4\pi}{3}\left(\frac{C\beta}{\pi f_a}\frac{B_0 m_a \tilde{\phi}}{8}\pi^2R^2\right)^2\left[
\frac{\tanh\left(\frac{\pi\kappa_{m_a+\Omega}R}{2}\right)^2}{\cosh\left(\frac{\pi\kappa_{m_a+\Omega}R}{2}\right)^2}+\frac{\tanh\left(\frac{\pi\kappa_{m_a-\Omega}R}{2}\right)^2}{\cosh\left(\frac{\pi\kappa_{m_a-\Omega}R}{2}\right)^2}
\right],\nonumber\\
\label{power4}
\eeq 
where $\kappa_{\omega}=\sqrt{\omega^2-\omega_P^2}$. Again, the time average of the instantaneous power is taken over a time scale larger than the longer time period, $\frac{2\pi}{|m_a-\Omega|}$. As expected, the condition for resonant radiation depends on both the frequency of the magnetic field and the plasma frequency. The radiation peaks for $R\sim \frac{1}{\kappa_{m_a+\Omega}}$ and for $R\sim\frac{1}{\kappa_{m_a-\Omega}}$. For $m_a\sim\Omega\sim\omega_P$, the first term in Eq. \ref{power4} peaks for radius $R\sim\frac{1}{\kappa_{m_a+\Omega}}\sim\frac{1}{m_a}$. If we now have $\kappa_{m_a-\Omega}\ll m_a\sim\Omega$, the second term in Eq. \ref{power4} peaks for condensate sizes much larger than $\frac{1}{m_a}$. If the condition, $\kappa_{m_a-\Omega}\ll m_a\sim\Omega$ is not satisfied, the second term in Eq. \ref{power4} is maximized for condensates of size $\frac{1}{m_a}$ as well.

\begin{figure}
	\centering
	\includegraphics{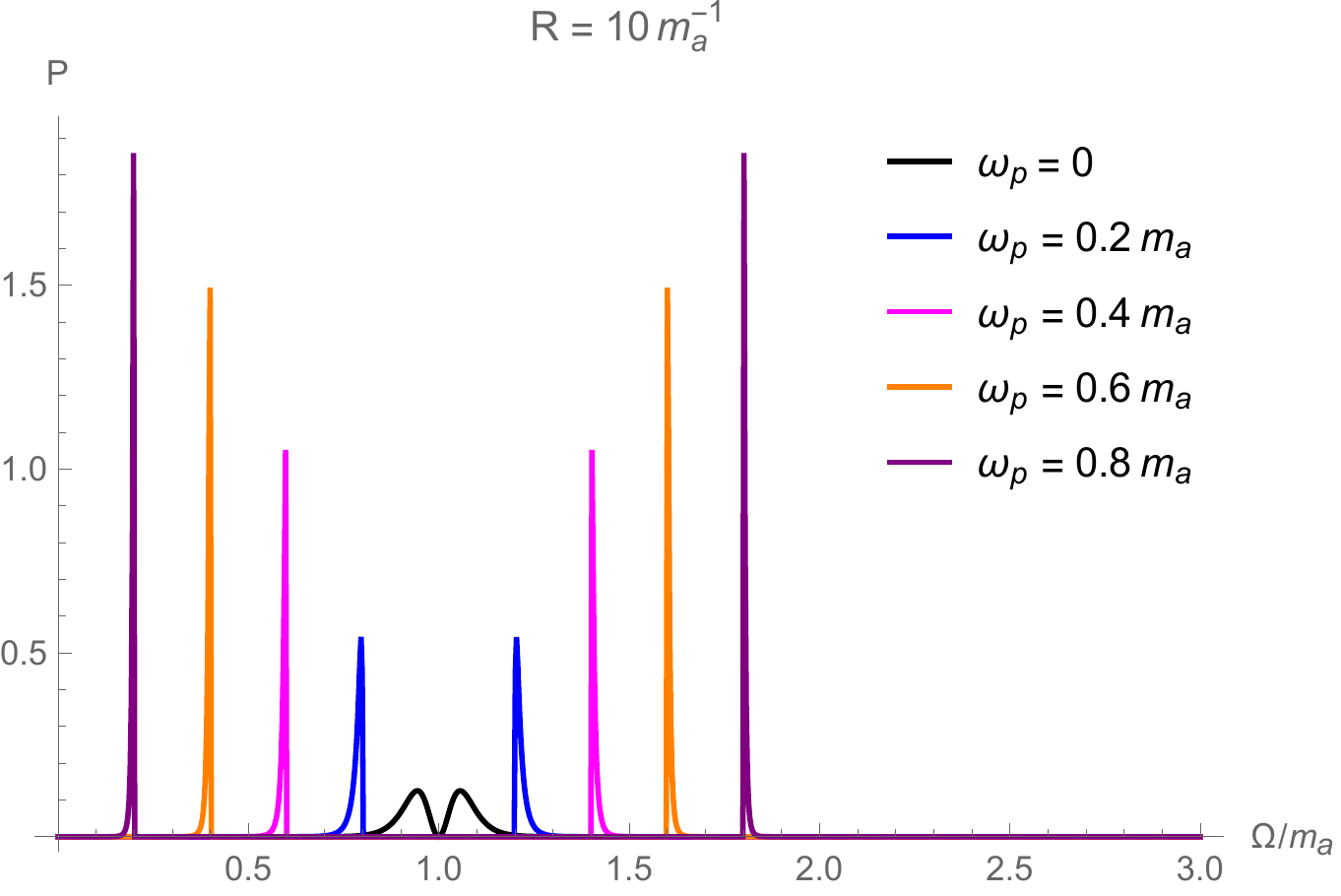}
	\caption{Radiated power as a function of a constant magnetic field frequency $\Omega$ for the case $R = 10m_a^{-1},$ averaged over 500 periods of the axion condensate. The power is measured in units of $\frac{4\pi}{3}\big(\frac{C\beta B_0 m_a \tilde{\phi}\pi^2 R^2}{8\pi f_a }\big)^2$.}\label{radiation_1}
\end{figure}

\begin{figure}
	\centering
	\includegraphics{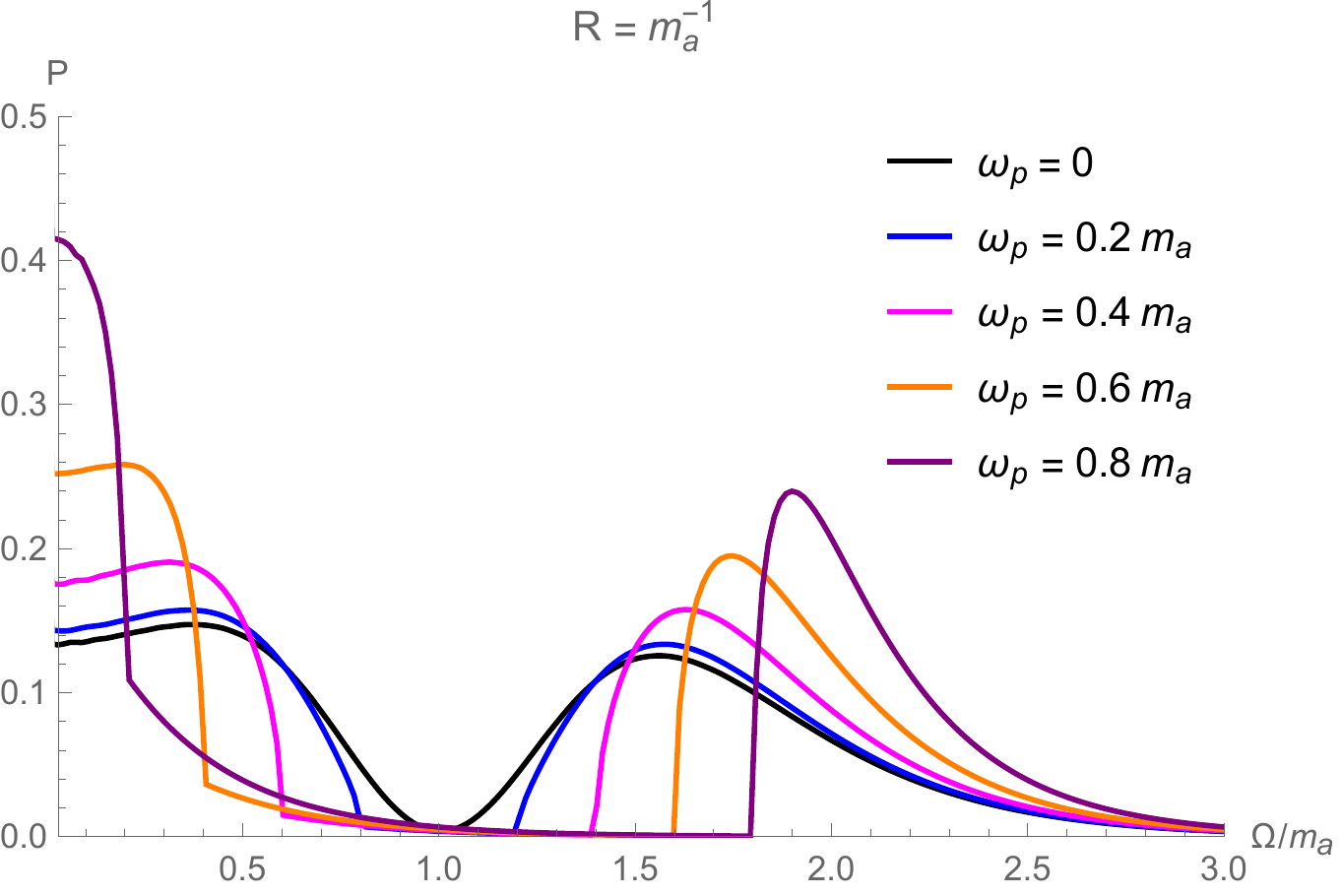}
	\caption{Radiated power as a function of a constant magnetic field frequency $\Omega$ for the case $R = m_a^{-1},$ averaged over 500 periods of the axion condensate. The power is measured in units of $\frac{4\pi}{3}\big(\frac{C\beta B_0 m_a \tilde{\phi}\pi^2 R^2}{8\pi f_a }\big)^2$.}\label{radiation_2}
\end{figure}

\begin{figure}
	\centering
	\includegraphics[width=.7\textwidth]{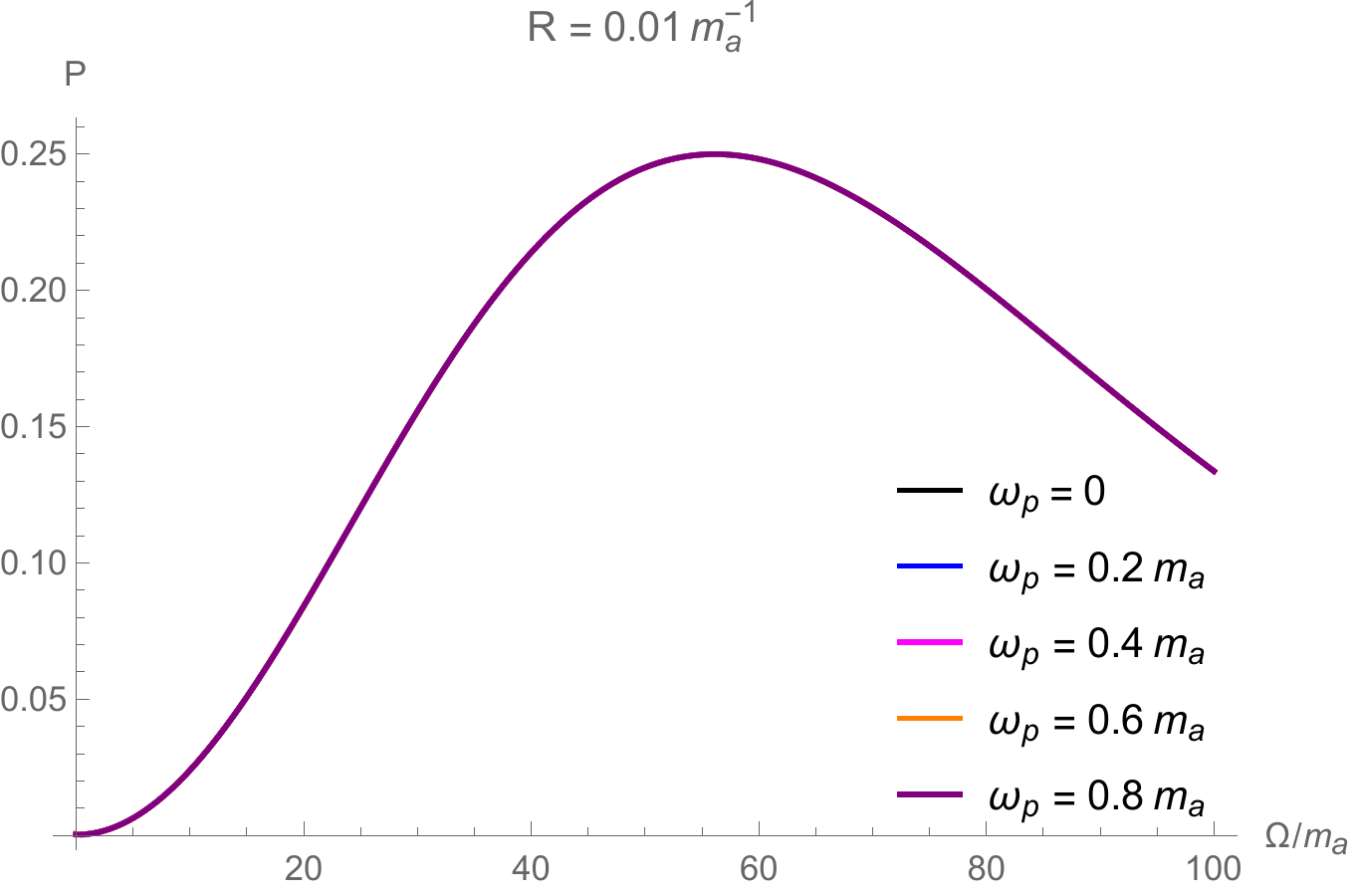}
	\caption{Radiated power as a function of a constant magnetic field frequency $\Omega$ for the case $R = 0.01m_a^{-1},$ averaged over 500 periods of the axion condensate. The power is measured in units of $\frac{4\pi}{3}\big(\frac{C\beta B_0 m_a \tilde{\phi}\pi^2 R^2}{8\pi f_a }\big)^2$.}\label{radiation_3}
\end{figure}

In figures \ref{radiation_1}-\ref{radiation_3} we show the radiated power as a function of $\Omega/m_a$ for the three cases $R=10 m_a^{-1}$, $R=m_a^{-1}$ and $R=0.01 m_a^{-1}$, respectively, for several different values of the plasma frequency. The power is averaged over several periods of oscillation for the axion condensate $\frac{2\pi}{m_a}$. % $500$ periods of the axion condensate, which for an axion mass of $\sim 10^{-12}$eV is on the order of seconds.
Our goal here is to understand how the radiated power changes with the magnetic field frequency $\Omega$ for various sizes of axion condensate. Let us first discuss the figures in Fig. \ref{radiation_1} and \ref{radiation_2}. The values of plasma frequency that we have considered are smaller than the axion mass scale. For these values of the plasma frequency, axion condensates of size $R=m_a^{-1}$ and $R=10 m_a^{-1}$, never realize the resonance condition $\kappa_{m_a+\Omega}=\frac{2(\log(\sqrt{2}+1))}{\pi R}$. In other words, the first term in square brackets in Eq. \ref{power4} does not reach its peak for any value of $\Omega>0$. However, the second term is maximized for $\kappa_{m_a-\Omega}=\sqrt{(\Omega-m_a)^2-\omega_P^2}=\frac{2(\log(\sqrt{2}+1))}{\pi R}$ and the corresponding resonance is observed for two different values of $\Omega$ with $\frac{\Omega}{m_a}=1\pm\sqrt{\left(\frac{2(\log(\sqrt{2}+1))}{\pi R m_a}\right)^2+\frac{\omega_P^2}{m_a^2}}$. Note that, each of these two peaks contain an abrupt drop in radiation on one side. This drop corresponds to the fact that for $1-\frac{\omega_P}{m_a}<\frac{\Omega}{m_a}<1+\frac{\omega_P}{m_a}$, there is no electromagnetic radiation as the frequency of radiation is below the photon's plasma mass. %The resonant peaks in Fig. \ref{radiation_1}, for $R\sim 10 m_a^{-1}$ are approximately located at $\Omega=m_a\pm \omega_P$ since $\frac{1}{(m_a R)^2}\ll \frac{\omega_P^2}{m_a^2}$. As a result, 
For $R\sim 10 m_a^{-1}$, the peaks are found symmetrically distributed around $\Omega=m_a$. 
This is because, for $R\sim 10 m_a^{-1}$ the first term in square brackets in Eq. \ref{power4} is negligible and the peaks observed in Fig. \ref{radiation_1} reflect the behavior of the second term in the square brackets. For $R\sim m_a^{-1}$ on the other hand,
the peaks are not as pronounced and they are not symmetrically distributed around $\Omega=m_a$. To understand this, note that although for $R=m_a^{-1}$, the first term in square brackets in Eq. \ref{power4} does not reach its peak for any $\Omega>0$, it is not completely negligible. Its tail contributes to Fig. \ref{radiation_2} accounting for the qualitative differences observed in the nature of the peaks located at $\Omega>m_a$ and $\Omega<m_a$ between Fig. \ref{radiation_1} and Fig. \ref{radiation_2}. We also note that as $R$ decreases, the peaks become wider: the peaks for $R\sim 10 m_a^{-1}$ are narrower than the ones for $R\sim m_a^{-1}$. To understand this, we have to remember that the peak width in Fig. \ref{radiation_1} and \ref{radiation_2} is controlled by the second term in the square brackets in Eq. \ref{power4}. The contribution to radiated power from this term is substantial only for $|\Omega-m_a|>\omega_P$ and $\sqrt{(\Omega-m_a)^2-\omega_P^2}<\frac{1}{R}$. The lower bound is set by the absence of propagating electromagnetic fields below the plasma frequency and the upper bound is set by the exponential drop off of radiation for wavelengths much larger than the radius of the condensate. The width of the peaks is thus $\delta\Omega\sim m_a\bigg(\sqrt{\left(\frac{1}{m_a R}\right)^2+\frac{\omega_P^2}{m_a^2}}-\frac{\omega_P}{m_a}\bigg)$. For $R m_a\gg \frac{m_a}{\omega_P}$, the width goes as $\delta \Omega\sim\frac{1}{\omega_P R^2}$, whereas for $Rm_a\sim\frac{m_a}{\omega_P}$, the peak width scales as $\delta\Omega\sim \omega_P$. Thus, the peaks in Fig.\ref{radiation_1} are much narrower than the peaks in Fig. \ref{radiation_2}. 

In the case of $R=0.01 m_a^{-1}$ shown in Fig. \ref{radiation_3}, the resonance condition is realized for both terms in the square bracket in Eq. \ref{power4} around $\Omega\sim\frac{2(\log(\sqrt{2}+1))}{\pi R}\gg m_a, \omega_P$. Here, the effect of the plasma frequency on the peaks is negligible and thus, the plots for different plasma frequencies overlap. The width of the peaks is set by the scale $\delta\Omega\sim\frac{1}{R}$ making them much wider than those in Fig. \ref{radiation_1} and \ref{radiation_2}. 

\begin{figure}
	\centering
	\includegraphics[width=.7\textwidth]{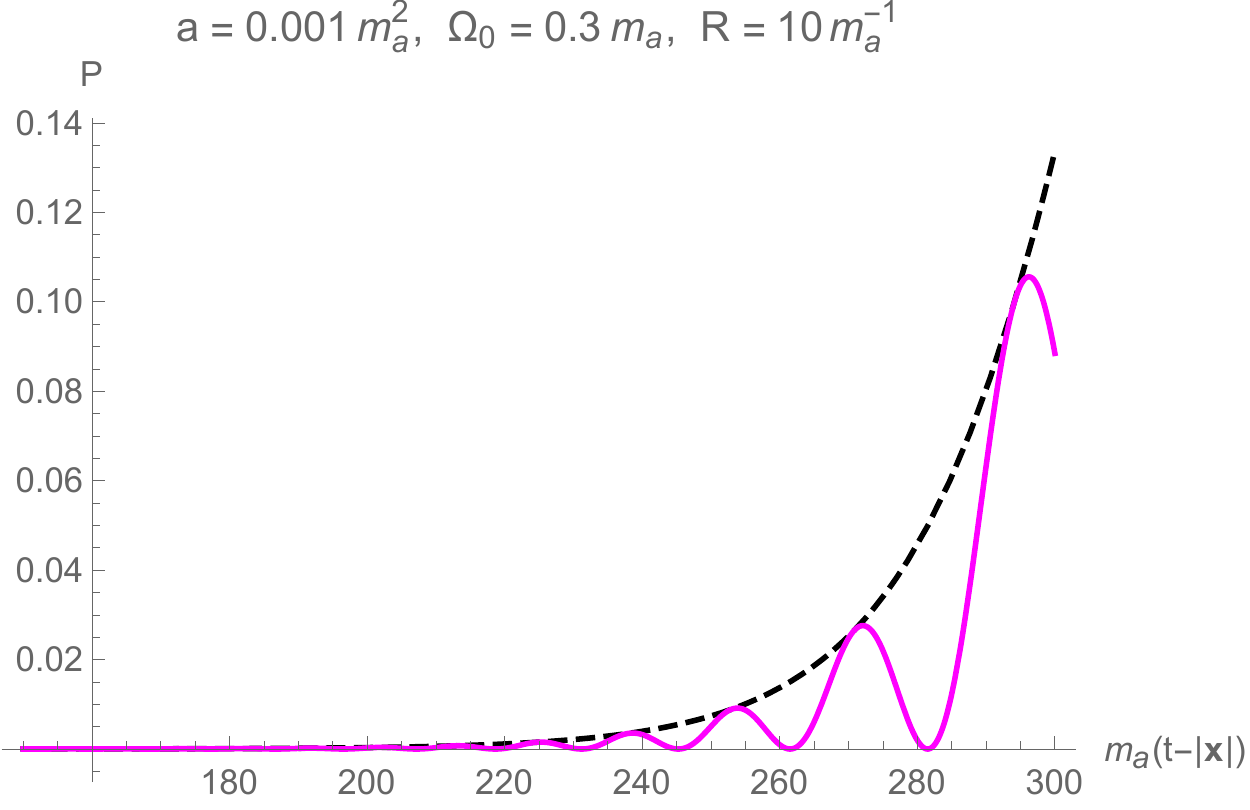}
	\caption{Instantaneous radiated power in the absence of a medium as a function of the retarded time $(t-\norm{\mathbf{x}})$ measured in units of the inverse axion mass for a linearly increasing magnetic field, $\Omega(t) = \Omega_0+a t$ (pink line), together with its corresponding envelope (black dashed line). We have here set $a = 0.001m_a^2$, $\Omega_0 = 0.3m_a$ and $R = 10 m_a^{-1}$. The power is measured in units of $\frac{4\pi}{3}\big(\frac{C\beta B_0 m_a \tilde{\phi}\pi^2 R^2}{8\pi f_a }\big)^2$.}\label{time_dependent}
\end{figure}
\subsubsection{Dissipation}
Previously while working with the frequency dependence of the conductivity, we took the collision-less limit $\omega\tau\gg 1$.
This resulted in a simple dispersion relation between the photon frequency and its wavelength $k=\sqrt{\omega^2-\omega_P^2}$ where we find propagating electromagnetic waves for $\omega>\omega_P$ and damped waves for $\omega<\omega_P$. As we will see, this conclusion is perfectly valid as long as $\sqrt{\omega^2-\omega_P^2}$ is not much smaller than $\omega_P$. When $\omega^2-\omega_P^2\ll \omega_P^2$, we have to be more careful. 
In order to understand why this is the case, note that in the presence of a frequency dependent conductivity in the collision-less limit we can write the dispersion relation for the photon as
\beq
k^2=\omega^2-\omega_P^2-i\frac{\omega_P^2}{\omega\tau}+\cdots.
\label{disp}
\eeq
We had previously ignored the third term on the RHS of Eq. \ref{disp}, in the limit of $\frac{1}{\omega\tau}\ll 1$. However, for $\omega\sim\omega_P$ we may not be able to ignore this term if $\frac{\omega^2-\omega_P^2}{\omega_P^2}\sim \frac{1}{\omega\tau}$. If $\frac{\omega^2-\omega_P^2}{\omega_P^2}\sim \frac{1}{\omega\tau}$, the wave vector has a non-negligible imaginary part and as a result the electromagnetic wave of frequency $\omega$ experiences significant damping. Since we found that the resonance in radiation can take place for long wavelengths with wave vectors $k_{m_a-\Omega}$ where $k_{m_a-\Omega}^{2}=(m_a-\Omega)^2-\omega_P^2\ll\omega_P^2\sim m_a^2$, we need to examine whether such wavelengths experience any significant damping more carefully. In order for resonance condition to be satisfied, and for radiation to peak for a frequency $\omega$ we know that $\omega^2-\omega_P^2\sim \frac{1}{R^2}$ and that $R\gg\frac{1}{m_a}\sim\frac{1}{\omega_P}$. However, for this wave to propagate without significant damping we will also need the condition $\omega^2-\omega_P^2\sim\frac{1}{R^2}\gg\frac{\omega_P^2}{\omega\tau}\sim \frac{\omega_P}{\tau}\implies \tau\gg \omega_P R^2$. So, in order for condensates of size much larger than $\frac{1}{m_a}, \frac{1}{\omega_P}$ for $\frac{1}{m_a}\sim\frac{1}{\omega_P}$ to radiate efficiently we need a scale separation between $\tau$ and $\frac{1}{m_a}$ such that
\beq
\tau\gg \omega_P R^2\sim\frac{\omega_P}{(m_a-\Omega)^2-\omega_P^2}\gg \frac{1}{m_a}.
\label{cond1}
\eeq 
In section \ref{ISM} we will examine to what extent such a condition is met in the interstellar medium. This will of course depend on the mass of axions we are interested in. The scale separation in question will be difficult to achieve if we lower the axion mass sufficiently such that the collision frequency of electrons becomes comparable with the axion mass scale. However, we will find that for the most interesting axion mass scales, this condition is met rather comfortably. 
\subsection{Time dependent frequency of alternating magnetic field}
At this stage we will attempt to understand the effect of a time dependent magnetic field frequency for the external magnetic field on radiation from axion condensates. We will simplify calculations by setting $\omega_P=0$ which amounts to neglecting the effects of a plasma. Incorporating a time dependence in the magnetic field frequency while also taking into account a finite plasma frequency for the photon is a rather involved calculation and we reserve this analysis for future work. As mentioned before the time dependence of the frequency of the magnetic field can be very complicated. However, our goal in this section is to demonstrate that in the presence of a time dependent external magnetic field frequency, it is possible for a condensate that is not radiating at an instant in time $t$ to radiate resonantly at another instant $t+\Delta t$. For the purpose of this demonstration, we don't need to consider the most general form for the frequency as a function of time. 
We will instead consider a linearly increasing magnetic field frequency $f(t)$ such that $\frac{d}{dt^n}f=0$ for $n>1$ and $\frac{df}{dt}\neq 0$. We will restrict our calculations to a regime where the change in the external magnetic field frequency over a single time period of the outgoing radiation is small compared to the frequency of the outgoing radiation. 

As an example, we begin with an external magnetic field of the form
\begin{equation}
\B(\x,t) =  B_0\cos(\Omega(t)t)\rz,
\end{equation}
with 
\begin{equation}
\Omega(t) = \Omega_0 +a t,
\end{equation}
where $\Omega_0$ is the frequency at time $t=0$.
The effective frequency of this time varying magnetic field is given by $\frac{d}{dt}(\Omega(t)t)$ which in this case is $\Omega_0+2at$. We will assume that $|(m_a-\Omega_0)|R\gg 1$ such that there is no significant radiation at any time $t$ when $a=0$. We will also assume $m_a>\Omega_0>0$ for convenience.  For an axion condensate of the form $\phi=\tilde{\phi}\cos(m_a t)\sech\left(\frac{r}{R}\right)$, the current density is given by
\begin{align}
\J(\x,t) &= -\g\big[(\pt\phi(\x,t))\B(t)\big]
\nonumber
\\
&= \g\frac{\tilde{\phi} m_a B_0}{2}\sech(\norm{\x}/R)\Big\{\sin\big[(m_a-\Omega(t))t\big]+\sin\big[(m_a+\Omega(t))t\big]\Big\}\rz.
\end{align}
Using the retarded Green's function the vector potential can then be expressed as

\begin{align}
\A(\x,t) &= -\int\dd^3x'\dd t' \frac{\delta(t-t'-\norm{\x-\x'})}{4\pi\norm{\x-\x'}}\g\frac{\tilde{\phi} m_a B_0}{2}\sech(\norm{\x'}/R)
\nonumber
\\
&\quad\quad\quad\times\Big\{\sin\big[(m_a-\Omega(t'))t'\big]+\sin\big[(m_a+\Omega(t'))t'\big]\Big\}\rz
\nonumber
\\
\nonumber
\\
&= -\g\frac{\tilde{\phi} m_a B_0}{8\pi}\int2\pi\,\dd\norm{\x'}\,\dd(\cos\theta')\norm{\x'}^2\frac{\sech(\norm{\x'}/R)}{\norm{\x-\x'}}
\nonumber
\\
&\quad\quad\quad\times\Big\{\sin\big[(m_a-\Omega_0)(t-\norm{\x-\x'})-a(t-\norm{\x-\x'})^2\big]
\nonumber
\\
&\quad\quad\quad\quad+\sin\big[(m_a+\Omega_0)(t-\norm{\x-\x'})+a(t-\norm{\x-\x'})^2\big]\Big\}\rz.
\end{align}
For $a=0$, the vector potential simplifies significantly in the radiation zone $|\x|\gg |\x'|$. However, with a time dependent frequency 
$a\neq 0$, the expression for the vector potential is somewhat more complicated and is given by 
\begin{align}
&\A(\x,t)
\nonumber\\ &\approx -\g\frac{\phi_0 m_a B_0}{4\norm{\x}}\int \dd\norm{\x'}\dd(\cos\theta')\norm{\x'}^2\sech(\norm{\x'}/R)
\nonumber
\\
&\quad\times\Big\{\sin\big[(m_a-\Omega_0)(t-\norm{\x}+\norm{\x'}\cos\theta')-a(t-\norm{\x})^2-2a(t-\norm{\x})\norm{\x'}\cos\theta'-a\norm{\x'}^2(\cos\theta')^2\big]
\nonumber
\\
&\quad+\sin\big[(m_a+\Omega_0)(t-\norm{\x}+\norm{\x'}\cos\theta')+a(t-\norm{\x})^2+2a(t-\norm{\x})\norm{\x'}\cos\theta'+a\norm{\x'}^2(\cos\theta')^2\big]\Big\}\rz. 
\label{A2}
\end{align}
Here $\theta'$ is the angle between $\x$ and $\x'$. At this point to make further progress we will assume $(t-|\x|)\gg R$ where $R$ is the size of the axion condensate. This allows us to ignore $\norm{\x'}^2(\cos\theta')^2$ in Eq. \ref{A2} and compute the integral in $\x'$ to find
\begin{align}
\A(\x,t) 
&=-\g\frac{\tilde{\phi} m_a B_0 R^2\pi^2}{8\norm{\x}}
\nonumber
\\
&\quad\quad\quad\times\bigg\{\frac{\tanh\big[\frac{\pi R}{2}\big((m_a-\Omega_0)-2a(t-\norm{\x})\big)\big]}{\cosh\big[\frac{\pi R}{2}\big((m_a-\Omega_0)-2a(t-\norm{\x})\big)\big]}\frac{\sin\big[(m_a-\Omega_0)(t-\norm{\x})-a(t-\norm{\x})^2\big]}{m_a-\Omega_0-2a(t-\norm{\x})}
\nonumber
\\
&\quad\quad\quad+\frac{\tanh\big[\frac{\pi R}{2}\big((m_a+\Omega_0)+2a(t-\norm{\x})\big)\big]}{\cosh\big[\frac{\pi R}{2}\big((m_a+\Omega_0)+2a(t-\norm{\x})\big)\big]}\frac{\sin\big[(m_a+\Omega_0)(t-\norm{\x})+a(t-\norm{\x})^2\big]}{m_a+\Omega_0+2a(t-\norm{\x})}\bigg\}\rz. 
\label{A3}
\end{align}
We will justify the validity of considering $(t-|\x|)\gg R$ at a later stage in the calculation.
We can now write the electric field in terms of the following variables 
\begin{align}
k^{(1)}_\pm = m_a\pm\big[\Omega_0+  a (t-\norm{\x})\big],
\\
k^{(2)}_\pm = m_a\pm\big[\Omega_0+ 2 a (t-\norm{\x})\big],
\label{def}
\end{align}
 as
\begin{align}
&\E_\text{r}'(\x,t)\nonumber\\ &=- \g\frac{\phi_0 m_a B_0 R^2\pi^2}{8\norm{\x}}\frac{\sech\big(\frac{\pi R}{2}k_-^{(2)}\big)}{(k_-^{(2)})^2}
\nonumber
\\
&\times\bigg\{a\pi R k_-^{(2)}\sech^2\big[\frac{\pi R}{2}k_-^{(2)}\big]\sin\big[(t-\norm{\x})k_-^{(1)}\big]
-(k_-^{(2)})^2\tanh\big[\frac{\pi R}{2}k_-^{(2)}\big]\cos\big[(t-\norm{\x})k_-^{(1)}\big]
\nonumber
\\
&\quad+2a\tanh\big[\frac{\pi R}{2}k_-^{(2)}\big]\sin\big[(t-\norm{\x})k_-^{(1)}\big]-a\pi R k_-^{(2)}\tanh^2\big[\frac{\pi R}{2}k_-^{(2)}\big]
\sin\big[(t-\norm{\x})k_-^{(1)}\big]\bigg\}\rz
\nonumber
\\
&\quad+(a\leftrightarrow-a, \Omega_0\leftrightarrow-\Omega_0). 
\end{align}
 In the limit of $a\ll( k_-^{(2)})^2$ and $a R\ll k_-^{(2)}$ the electric field can be simplified further 
\begin{align}
\E_\text{r}'(\x,t) &\approx \g\frac{\phi_0 m_a B_0 R^2\pi^2}{8\norm{\x}}.
\nonumber
\\
&\quad\quad\quad\times\bigg\{\frac{\tanh(\pi R k_-^{(2)}/2)}{\cosh(\pi R k_-^{(2)}/2)}\cos\big[(t-\norm{\x})k_-^{(1)}\big]+\frac{\tanh(\pi R k_+^{(2)}/2)}{\cosh(\pi R k_+^{(2)}/2)}\cos\big[(t-\norm{\x})k_+^{(1)}\big]\bigg\}\rz
\label{ez2}
\end{align}
Similarly, the magnetic field is given by
\begin{equation}
\B_\text{r}(\x,t) = \norm{\E_\text{r}(\x,t)}(\hat{\x}\times\rz).
\end{equation}
We now arrive at the expression for the instantaneous radiated power
\begin{align}
P_i &= \frac{4\pi}{3}\g^2\bigg(\frac{\tilde{\phi} m_a B_0 R^2\pi^2}{8}\bigg)^2
\nonumber
\\
&\quad\quad\quad\times\bigg\{\bigg(\frac{\tanh(\pi R k_-^{(2)}/2)}{\cosh(\pi R k_-^{(2)}/2)}\bigg)^2\cos^2\big[(t-\norm{\x})k_-^{(1)}\big]+\bigg(\frac{\tanh(\pi R k_+^{(2)}/2)}{\cosh(\pi R k_+^{(2)}/2)}\bigg)^2\cos^2\big[(t-\norm{\x})k_+^{(1)}\big]
\nonumber
\\
&\quad\quad\quad\quad+2\frac{\tanh(\pi R k_-^{(2)}/2)}{\cosh(\pi R k_-^{(2)}/2)}\frac{\tanh(\pi R k_+^{(2)}/2)}{\cosh(\pi R k_+^{(2)}/2)}\cos\big[(t-\norm{\x})k_-^{(1)}\big]\cos\big[(t-\norm{\x})k_+^{(1)}\big]\bigg\}.
\label{power5}
\end{align}
It is easy to see that the radiated power for an external magnetic field of constant frequency is recovered in the limit of $a=0$. 
Note that, even though the sinusoidal variation in the radiated power is expressed as a function of $k_{\pm}^{(1)}$, the effective frequency 
and wavelengths of these waves are set by $\frac{d}{d(t-|\x|)}\left(k_{\pm}^{(1)}(t-|\x|)\right)\sim k_{\pm}^{(2)}$. As is clear from Eq. \ref{power5}, for condensates of size $R\gg\frac{1}{m_a}\sim \frac{1}{k_+^{(2)}}$ a resonant 
enhancement in radiation can take place in the limit of $k_{-}^{(2)}\sim 2\frac{\log(\sqrt{2}+1)}{\pi R}$ when the first term inside the curly brackets in Eq. \ref{power5} dominates over the others which are exponentially suppressed. We can define a time averaged radiated power 
by averaging the instantaneous power over a few time slices of interval $\frac{1}{k_-^{(2)}}$ as long as the change in frequency over one time period is much smaller than the frequency itself, i.e. $\frac{2\pi a}{k_-^{(2)}}\ll k_-^{(2)}$:
\begin{align}
P_{\text{av}} &\approx\left(\frac{C\beta}{\pi f_a}\right)^2\frac{\tilde{\phi}^2B_0^2m_a k_{m_a}R^4\pi^5}{12}\bigg(\frac{\tanh(\pi R k_-^{(2)}/2)}{\cosh(\pi R k_-^{(2)}/2)}\bigg)^2.
\label{power61}
\end{align}
Near the resonant peak we also have $\pi R k_-^{(2)}/2\sim\log(\sqrt{2}+1)$. Thus, we have $a\ll (k_-^{(2)})^2\sim k_-^{(2)}/R$ where we have ignored order $\sim 1$ numerical factors. Note that these very limits were used to arrive at Eq \ref{ez2}. So, our result in Eq. \ref{power61} is self consistent. 
At this stage we can also check the validity of the condition $t-|\x|\gg R$ near the resonant peak. Recall that this condition was used to derive Eq. \ref{A3}. The resonant peak in radiation is at $k_{-}^{(2)}\sim 2\frac{\log(\sqrt{2}+1)}{\pi R}$. Using the definitions in Eq. \ref{def} we find 
\beq
t-|\x|&\sim& \frac{(m_a-\Omega_0)-\frac{2(\log(\sqrt{2}+1))}{\pi R}}{2a}\nonumber\\
&\sim &\frac{(m_a-\Omega_0)}{2a},\nonumber\\
\eeq
where in the last line we have taken the limit $(m_a-\Omega_0)\gg\frac{2(\log(\sqrt{2}+1))}{\pi R}$ for large condensates ($Rm_a\gg 1$) and $m_a-\Omega_0\sim m_a\sim\Omega_0$. Thus, we see that, the condition of $t-|\x|\gg R$ near the peaks translates to $m_a-\Omega_0\gg 2a R$. Our analytical calculation for the resonant peak for a time varying frequency therefore involves controlled approximations in the limit of $\frac{(m_a-\Omega_0)}{2a}\gg R\gg(m_a-\Omega_0)^{-1}$. We previously imposed the condition $a\ll (k_-^{(2)})^2$ which along with $R\sim\frac{1}{k_-^{(2)}}$ implies $aR^2\ll 1$. Defining $aR^2\equiv b$, we can rewrite the condition as $\frac{(m_a-\Omega_0)R}{2b}\gg 1\gg((m_a-\Omega_0)R)^{-1}$ which is clearly satisfied for any $(m_a-\Omega_0)R\sim m_aR\gg 1$.
Note that some of the limits we assumed were needed in order to perform a controlled analytic calculation. These limits can be relaxed 
if the analysis is performed numerically. 

In Fig. \ref{time_dependent} we demonstrate how a time varying frequency of the external alternating magnetic field can affect radiation from a large axion condensate, $R = 10m_a^{-1}$. In the figure we have set the rate of change of the magnetic field frequency to be $0.001m_a^2$, which for an axion mass of $\sim10^{-12}$eV corresponds to about $\sim 10^3\text{s}^{-2}$. We set the oscillating magnetic field frequency at $t=0$, $\Omega_0 = 0.3m_a$. The corresponding radiation reaches position $|\x|$ at time $t=|\x|$. However, this radiated power is negligible due to the large size of the condensate: $R\gg\frac{1}{k_\pm^{(2)}(t-\norm{\mathbf{x}} = 0)}$. We see that the radiated power increases with time as the frequency of the magnetic field increases and resonant conversion is eventually realized for $k_-^{(2)}\sim R^{-1}$. This shows that a wide range of axion masses can trigger resonant conversion, if the frequency of the magnetic field is dependent on time. 

\section{Properties of interstellar plasma and alternating magnetic fields of Neutron stars}
\label{ISM}
In this section we will discuss the time scales associated with some astrophysical sources of time dependent magnetic fields, e.g. pulsars and neutron stars. We will also discuss how the plasma scales relevant for the interstellar medium compare with these time scales.
We mentioned earlier that alternating magnetic fields can be sourced by spinning neutron stars. To understand this, note that the spin of a neutron star can range from a few Hz to a few kHz for pulsars\cite{ERA}. A time period of $10^{-3}$s translates to a energy scale of $10^{-12}$ eV in natural units. Since pulsars host very strong magnetic fields which are typically at an angle with the axis of rotation, they can be thought of as rotating magnetic dipoles. In fact, this rotating magnetic dipole configuration of pulsars is expected to cause spin down of the star by emitting magnetic dipole radiation. The frequency of this oscillating magnetic dipole is set by the spin frequency of the star(typically a few kHz) and interestingly it is close to the plasma frequency of the interstellar medium \cite{ERA}  
\beq
\nu_P&\sim&\left(\frac{e^2 n_e}{\pi m_e}\right)^{1/2}\nonumber\\
&\approx& 8.97\,\,\text{kHz}\left(\frac{n_e}{\text{cm}^{-3}}\right)^{1/2}.
\eeq
For typical interstellar medium (ISM), one can take $n_e\sim 0.03\text{cm}^{-3}$ which results in $\nu_P\sim 1.5 \,\,\text{kHz}$. \footnote{In fact, the presence of the interstellar plasma introduces significant damping of the magnetic dipole radiation which in turn causes interstellar medium to heat up.} 
Thus, an axion condensate in the vicinity of a pulsar can experience an alternating magnetic field, the frequency of which is of the order of the plasma frequency of the interstellar medium. If the axion mass for such a condensate is also of the same order, $\sim 10^{-12}$ eV, we can expect the axion condensate to decay through electromagnetic radiation. Axions of mass $10^{-12}$ eV are considered ultralight and can form axion stars depending on the details of their interactions. They can also form superradiant axion condensates around black holes of a few solar masses\cite{Arvanitaki:2009fg, Brito:2015oca, Arvanitaki:2014wva, Arvanitaki:2016qwi}. Thus, any time a rotating neutron star interacts with an axion star or a superradiant condensate of axion mass $10^{-12}$ eV, there can be significant electromagnetic radiation which can deplete the axion condensate and leave its imprints in gravitational waves.

Note that, it is not just pulsar spin which can produce alternating magnetic field of frequency a few Hz to kHz. Similar alternating fields can result from merging neutron stars which sweep a frequency of $10$ Hz to a kHz during inspiral \cite{Piro:2012rq}. This can give rise to alternating magnetic fields with increasing frequency in time which were discussed in the previous section. 

Let us now recall our discussion on dissipation from the subsection \ref{plasma} where we noted that for resonant radiation to take place, the condition in Eq. \ref{cond1} needs to be satisfied. We will now verify to what extent this condition is met for ultralight axion condensates in the ISM background and in an alternating external field of frequency a few kHz. The collision frequency of electrons in interstellar medium was discussed in \cite{Sen:2018cjt} $\sim 10^{-18}\left(\frac{1 \text{eV}}{T}\frac{n_e}{1/\text{cc}}\right)$. Taking the temperature to be $1$ eV and a density of $n_e\sim 0.01\text{cm}^{-3}$ we see that the inequality of Eq. \ref{cond1} is indeed satisfied for $\omega_P\sim m_a\sim 10^{-11}-10^{-12}$ eV. As an example, we can expect axion condensates of size $R\gg\frac{1}{m_a}$ for $m_a\sim 10^{-12}$ eV to radiate resonantly as long as $R^2<\frac{\tau}{\omega_P}\sim \frac{10^8}{m_a^2}$. 

\section{Axion decay time scale}
Since electromagnetic radiation takes away energy from the axion condensate, the condensate amplitude will decay with time.
Our analysis of electromagnetic radiation from axion condensates so far ignores the effect of back reaction or the decay of the condensate. As mentioned in the introduction, this is justified so long as the decay time scale of the condensate is larger than the time period of the outgoing electromagnetic waves.  
Here we will estimate this decay time scale taking into account resonant enhancement in radiation for condensates of size $R\sim\frac{1}{m_a}$ and large condensates $R\gg \frac{1}{m_a}$. From Eq. \ref{power4} we know that radiation can peak for axion condensates of size $R\sim\frac{1}{\kappa_{m_a+\Omega}}\sim \frac{1}{m_a}\sim\frac{1}{\Omega}$ owing to the first term on the RHS of Eq. \ref{power4}. Similarly, radiation can peak for condensates of size $R\sim\frac{1}{\kappa_{m_a-\Omega}}$ owing to the second term on the RHS. Note that the energy density stored in an axion condensate is of the order of $m_a^2\tilde{\phi}^2\sech\left(\frac{r}{R}\right)^2$ and the corresponding energy is $\sim m_a^2\tilde{\phi}^2\frac{\pi^3R^3}{3}$. So the decay time scale is given by 
\beq
T\sim\frac{\frac{\pi^3R^3}{3}m_a^2\tilde{\phi}^2}{\frac{4\pi}{3}\left(\frac{C\beta\tilde{\phi}}{\pi f_a}\frac{B_0 m_a}{8}\pi^2 R^2\right)^2\left[
\frac{\tanh\left(\frac{\pi\kappa_{m_a+\Omega}R}{2}\right)^2}{\cosh\left(\frac{\pi\kappa_{m_a+\Omega}R}{2}\right)^2}+\frac{\tanh\left(\frac{\pi\kappa_{m_a-\Omega}R}{2}\right)^2}{\cosh\left(\frac{\pi\kappa_{m_a-\Omega}R}{2}\right)^2}\label{time_scale}
\right]}.
\eeq
 For condensate of size $R\sim\frac{1}{\kappa_{m_a+\Omega}}$, if the scales $\kappa_{m_a+\Omega}$ and $\kappa_{m_a-\Omega}$ are very different, the decay time scale goes as
\beq
T\sim\frac{\pi^3R^3m_a^2\tilde{\phi}^2}{\frac{\pi}{16}\left(\frac{C\beta\tilde{\phi}}{\pi f_a}B_0 m_a\pi^2R^2\right)^2}\sim \frac{16}{m_a\pi^2\left(\frac{C\beta}{\pi m_af_a}B_0\right)^2}\left(\frac{\kappa_{m_a+\Omega}}{m_a}\right).
\label{T1}
\eeq
Similarly, for a condensate of size $R\sim\frac{1}{\kappa_{m_a-\Omega}}$, the decay time scale goes as
\beq
T\sim\frac{\pi^3R^3m_a^2\tilde{\phi}^2}{\frac{\pi}{16}\left(\frac{C\beta\tilde{\phi}}{\pi f_a}B_0 m_a\pi^2R^2\right)^2}\sim \frac{16}{m_a\pi^2\left(\frac{C\beta}{\pi m_af_a}B_0\right)^2}\left(\frac{\kappa_{m_a-\Omega}}{m_a}\right).
\label{T2}
\eeq
The two time scales obtained in Eq. \ref{T1} and \ref{T2} can help us identify the region in parameter space where back reaction of axion condensates can be ignored. In order to ignore back reaction, for $R\sim\frac{1}{\kappa_{m_a+\Omega}}$, the time scale of decay given in Eq. \ref{T1} should be larger than the time period of outgoing radiation given by $\frac{2\pi}{|m_a+\Omega|}\sim \frac{2\pi}{m_a}$. We can rewrite the time scale in Eq. \ref{T1} substituting $\kappa_{m_a+\Omega}\sim m_a\sim\Omega\sim\omega_P$ as
\beq
T\sim \frac{16}{m_a\pi^2\left(\frac{C\beta}{\pi m_af_a}B_0\right)^2}.
\label{T11}
\eeq
This time scale is larger than $\frac{2\pi}{m_a}$ so long as $\left(\frac{C\beta}{\pi m_af_a}B_0\right)^2$ is not much larger than $1$. In the case of QCD axions, $m_a f_a\sim \Lambda_{\text{QCD}}^2$ and even for the strongest fields in pulsars we can take $B_0< \Lambda_{\text{QCD}}^2$. As an example, a magnetic field of $10^{14}$ Gauss found in magnetars translates to an energy scale of $\sim 1\, \text{MeV}^2$.
As a result, at weak coupling, the time scale in Eq. \ref{T11} is much larger than the time period of outgoing radiation. 

When $R\sim\frac{1}{k_{m_a-\Omega}}$, the relevant decay time scale of Eq. \ref{T2} must be larger than the time period of the outgoing radiation given by $\frac{2\pi}{|m_a-\Omega|}$. For $\kappa_{m_a-\Omega}\ll m_a$ and $m_a\sim\Omega\sim\omega_P$ the time period of the outgoing wave $\frac{2\pi}{|m_a-\Omega|}\sim \frac{2\pi}{m_a}$. Thus, in order to ignore back reaction we need
\beq
T\sim \frac{16}{m_a\pi^2\left(\frac{C\beta}{\pi m_af_a}B_0\right)^2}\left(\frac{\kappa_{m_a-\Omega}}{m_a}\right)> \frac{2\pi}{m_a}.
\label{T22}
\eeq
This condition is satisfied as long as $\left(\frac{C\beta}{\pi m_af_a}B_0\right)^2\leq \left(\frac{\kappa_{m_a-\Omega}}{m_a}\right)\ll 1$.
If we set $\left(\frac{C\beta}{\pi m_af_a}B_0\right)^2\sim \left(\frac{\kappa_{m_a-\Omega}}{m_a}\right)$ the decay time scale is of the order $\frac{1}{m_a}$ which can be a few seconds or smaller for ultralight axions. 

We conclude this section with an estimate for the total energy released in electromagnetic radiation. As discussed in \cite{Visinelli:2017ooc}, for moderately dense axion condensates, $R\sim m_a^{-1}$ and $\tilde{\phi}\sim f_a$. In this case, we can write the total radiated energy as $m_a^2 f_a^2 R^3\sim f_a^2/m_a$. For an axion mass of $10^{-10}-10^{-12}$ eV, the energy released is of the order of a solar mass. Large condensates $R\gg m_a^{-1}$, which are also referred to as dilute condensates, on the other hand can release energy of the order of $\frac{m_P^2}{m_a}\frac{1}{(m_a R)^2}$, where $m_\text{P}$ is the Planck mass \cite{Visinelli:2017ooc}. For an axion mass of $10^{-10}-10^{-12}$ eV and $R m_a\sim 10-100$, this estimate too can be of the order of a few solar masses. Thus we see that large axion condensates(dilute) $Rm_a\gg 1$ can radiate as efficiently as smaller ones $R\sim m_a^{-1}$ (dense) if there is resonant enhancement in radiation for the former. In both cases, energy of the order of a few solar masses can be radiated away within a time scale of seconds.

\iffalse
\hl{We end off this section by estimating the average power radiated near the resonance for large condensates. In \cite{Visinelli:2017ooc} it is found that the amplitude of a dilute axion condensate ($R\gg m_a^{-1}$) in the absence of self interactions, is given by $\tilde{\phi}\sim\frac{f_a m_\text{pl}}{m_a\Lambda^2 R^2}$,} \textcolor{red}{(In order to find this, combine these five equations from section B of that paper: $\tilde{M} = M\frac{m_a}{f_a^2}$, $\tilde{R} = m_a R$, $\beta = (\frac{m_\text{pl}}{f_a})^2$, $\Theta_0^2\sim \frac{M}{\Lambda^4 R^3}$, $\tilde{R}\sim\frac{1}{\beta \tilde{M}}$, together with the definition $\tilde{\phi} = f_a \Theta_0$)} \hl{where $m_{\text{pl}}\sim 10^{19}$GeV is the planck mass, and $\Lambda \sim 75$MeV. Using that $m_a f_a\sim \Lambda_\text{QCD}^2\sim \Lambda^2$, we then have 
$\tilde{\phi}\sim \frac{10^{19}\text{GeV}}{(m_a R)^2}$. For large $R$, the resonant radiation only occurs for the second term in} \eqref{power4}, \hl{when $R\sim k_{m_a-\Omega}^{-1}$. In that case we can estimate the average power radiated to be:}
\beq
P_\text{av}&\sim&\frac{4\pi}{3}(10^{38}\text{GeV}^2)\left(\frac{C\beta B_0}{m_a\pi f_a}\frac{\pi^2}{8}\right)^2.
\eeq 
\hl{The constant $\frac{C\beta}{\pi}$ is of order 1, and thus for a magnetic field of size $B_0\sim 10^{14}$gauss, we are then left with }
\beq
P_\text{av}&\sim&(10^{38}\text{GeV}^2)\left(10^{-8}\right)\sim10^{30}\text{GeV}^2. 
\eeq 
\hl{This corresponds roughly to radiating away $\sim10^{-3}$ solar masses every second. }
\fi
\section{conclusion}
Time dependent magnetic fields are ubiquitous in astrophysics. Strong time dependent fields are often found in pulsars, supernovae and mergers of neutron stars. We can expect electromagnetic radiation from axion condensates in the background of such external magnetic fields to experience significant modification due to this time dependence, provided the corresponding frequency scale is comparable to the axion mass scale. In this paper we demonstrate this effect by considering the most simple form of time dependence for the external magnetic field: a sinusoidal variation in time. We choose such a form  for the time dependence in order to extract the physics of interest which clarifies the effect of the two competing scales on radiation. We reserve the analysis of a more realistic time dependence for future work. Our work however captures the essential features of radiation from axion condensate in a time dependent magnetic field. 
We find that, large axion condensates $Rm_a\gg 1$, which don't radiate efficiently in general, can do so in the presence of an oscillating background field provided the frequency of oscillation is close to the axion mass scale. This enhancement in radiation is somewhat analogous to the resonant enhancement in radiation of axion condensates experienced in a plasma background. We also analyzed the combined effects of an oscillating magnetic field and a plasma medium on axion radiation. Not surprisingly, we find that the condition for resonant enhancement of radiation now depends on both the plasma frequency and the background field frequency. This enhancement in radiation can result in a release of few solar mass worth energy in a few seconds for ultralight QCD axion of mass $10^{-10}-10^{-12}$ eV. Of course, this enormous release of energy can also take place for ultralight axion like particles. We also emphasize that several astrophysical phenomena can give rise to alternating magnetic fields with a frequency which itself varies in time. Such time varying frequencies can result in resonant radiation for a range of axion mass scales at a certain interval in time when the two frequency scales (the mass of the axion and the frequncy of the time dependent external field) become comparable. This can enable a non-radiating axion condensate at a certain instant in time to radiate efficiently at a later time. We demonstrate this behavior with a magnetic field frequency varying linearly in time in the absence of a plasma background. Again, the physics of interest is not dependent on the specific form of time dependence that the magnetic field frequency can have. Our analysis can easily be extended to any other slow variation for the time dependent frequency, as long as the change in magnetic field frequency over a single time period of the outgoing radiation is smaller than the frequency of the outgoing radiation. In future work we plan to consider a more general form for the time dependence including monotonic growth in the magnetic field. Such magnetic fields can be found in supernovae and post merger dynamics of neutron stars \cite{Kiuchi:2015sga, Skoutnev:2021chg}. We also intend to analyze the effects of spatial variation in background fields and the plasma frequency for large axion condensates in future work.   
\section{acknowledgment}
This research was supported in part by DOE grant number GR-024204-00001.
\bibliographystyle{utphys}
\bibliography{ref}

 \end{document}